\documentstyle[11pt]{article}
\newlength{\dinwidth}
\newlength{\dinmargin}
\newcommand{\ga}{\alpha}
\newcommand{\gb}{\beta}
\newcommand{\gc}{\gamma}
\newcommand{\gC}{\Gamma}
\newcommand{\gd}{\delta}

\newcommand{\gep}{\epsilon}

\newcommand{\gT}{\Theta}

\newcommand{\gy}{\eta}

\newcommand{\di}{\partial}

\newcommand{\bgep}{\bar{\epsilon}}

\newcommand{\uba}{{\underline{a}}}
\newcommand{\ubb}{{\underline{b}}}
\newcommand{\ubc}{{\underline{c}}}

\newcommand{\uga}{{\underline{\ga}}}
\newcommand{\ugb}{{\underline{\gb}}}
\newcommand{\ugc}{{\underline{\gc}}}
\newcommand{\ugd}{{\underline{\gd}}}

\newcommand{\lb}[1]{\label{#1}}
\newcommand{\Eq}[1]{(\ref{#1})}

\renewcommand{\[}{\begin{eqnarray}}
\renewcommand{\]}{\end{eqnarray}}
\newcommand{\nn}{\nonumber}
\newcommand{\non}{\nonumber \\ }

\begin{document}
\arraycolsep3pt
\thispagestyle{empty}
\begin{flushright} hep-th/9912226\\
                   KCL-MTH-99-49
\end{flushright}
\vspace*{2cm}
\begin{center}
 {\LARGE \sc Brane Rotating Symmetries \\[1ex]
             and the Fivebrane Equations of Motion\\
\vspace*{1cm}}
 {\sl
     Oliver B\"arwald\footnote[1]{Supported by the EC under TMR
     contract ERBFMBICT972717,},
     and Peter C.\ West\\
oliver, pwest@mth.kcl.ac.uk\\
 \vspace*{6mm}
     Department of Mathematics, King's College London\\
     Strand, London WC2R 2LS, Great Britain\\
 \vspace*{6mm}}
{\today}\\
\vspace*{1cm}

\begin{minipage}{11cm}\footnotesize {\bf Abstract:} We show that the
fully covariant equations of motion for the M-theory fivebrane can be
interpreted as charge conservation equations.  The associated charges
induce `shift'-symmetries of the scalar, spinor and gauge-fields of
the fivebrane, so allowing an interpretation of all these fields as
Goldstone fields.  We also find that the fivebrane possesses a new
symmetry that is part of the $GL(32)$ automorphism group of the eleven
dimensional supersymmetry algebra.
\end{minipage}
\end{center}

\noindent In this paper we give a new interpretation to the fivebrane
equations of motion.  We show that these can be written as
conservation equations for three conserved currents.  The symmetries
generated by the associated charges correspond to constant `shifts' of
the scalar, spinor and gauge-fields respectively.

We then compute the algebra generated by these charges.  This turns
out to be -- as expected -- the eleven dimensional supersymmetry
algebra with two-form and five-form central charges present.  Hence in
this manner we have reconstructed the eleven dimensional supersymmetry
from the fivebrane equations of motion.  This superalgebra has
previously been derived \cite{SorTow97} in the context of the
Lagrangian formulation of the fivebrane.  By reducing the eleven
dimensional supersymmetry algebra, we find the (2,0) worldvolume
supersymmetry algebra.  The currents that give rise to this
worldvolume superalgebra are explicitly constructed.

We also find that the fivebrane contains a new symmetry that is part
of the $GL(32)$ automorphism group of the eleven dimensional
supersymmetry algebra.  This symmetry rotates the fivebrane and
twobrane into each other.  Finally, we discuss some of the possible
consequences for M theory of this new symmetry.

\section{Conserved Currents and the Equations of Motion}
The fivebrane-equations of motion for a flat background and ignoring
higher order terms in the fermion fields are given by \cite{HoSeWe97a}
\[
G^{mn} \nabla_{m} \nabla_n X^{\uba} &=&0\lb{scalarEOM},\\
\nabla^{m} \gT (1-\gC)\gC^{n} m_{mn}&=&0\lb{spinorEOM},\\
G^{mn} \nabla_{m} H_{npq} &=&0\lb{tensorEOM}.
\]
We will now define the symbols occurring in the equations of motion:
Our index-convention is to use $m,n,p,\ldots=0,1,\ldots,5$ and
$a,b,c,\ldots=0,1,\ldots,5$ for world-volume world and tangent indices
respectively.  For the flat target-space indices we use
$\uba,\ubb,\ubc,\ldots=0,1,\ldots ,10$.  In static gauge the
target-indices can be decomposed into world-volume and transverse
indices as $\uba= (a,a')$.  For fermionic indices it is customary to
use same scheme with Greek letters, however we are going to suppress
these indices in most of the following for simplicity.  We will use
the `mostly plus' convention for the Minkowski-metric $\eta_{ab} :=
{\rm diag}(-1,1,\ldots,1)$ and normalize the $\gep$-tensor as
$\gep^{012345}=1$.

The metric on the brane is the pullback of
the flat target space metric
\[
g_{mn} :=  \di_{m}X^{\uba} \di_{n}X^{\ubb}\gy_{\uba\ubb},
\]
and the covariant derivative is defined with the Levi-Civita
connection with respect to this metric. We define the
associated vielbein
in the standard way as
\[
e_{m}{}^a e_n{}^b \gy_{ab} := g_{mn}.
\]
It will be necessary to switch frequently between frames.
To make life
simpler we will therefore normally suppress the vielbein-factors.

The tensor $G^{mn}$ is
related to the induced metric via
\[
G^{mn}:= (m^{2})^{mn},\lb{bigG}
\]
where the matrix $m$ is given in tangent frame as
\[
m_{ab} := \eta_{ab} - 2 h_a{}^{cd} h_{bcd}\equiv \eta_{ab}-2 k_{ab}.
\]
The 3-form field $h_{abc}$ is self-dual
\[
h_{abc} = \frac{1}{3!} \gep_{abcdef}h^{def},
\]
but it is not the curl of a two-form gauge field. It is related to
the field $H_{mnp}=3\di_{[m} B_{np]}$ which appears in the equations
of motion, but $H_{mnp}$ is not self-dual in the linear sense.
The relationship between the two fields is given by
\[
H_{abc} = (m^{-1})_c{}^d h_{abd}.
\]
Finally the matrix $\gC$ appearing in the equation of motion for the
fermions is given by
\[
\gC = -\frac{1}{6!}\frac{1}{\sqrt{-g}} \gep^{m_1\ldots m_6}
\gC_{m_1 \ldots m_6} + \frac{1}{3} h^{mnp}\gC_{mnp}.
\]
It satisfies $\rm{Tr} \gC = 0$ and $\gC^{2} = 1$ and hence gives rise to
the projector $\frac12 (1-\gC)$, projecting onto half of the original
spinor space.

The covariantly conserved energy-momentum tensor for this system is
given by $T^{mn} = Q^{-1} G^{mn}$ \cite{BaLaWe99}
where $Q= 1- {2\over 3}k^{ab}k_{ab}$.  Using this we can
rewrite \Eq{scalarEOM} as
\[
0=T^{mn}\nabla_m \di_n X^{\uba}=\nabla_m(T^{mn}\di_n
X^{\uba})=\frac{1}{\sqrt{-g}} \di_{m}(\sqrt{-g}T^{mn}\di_n X^{\uba}).
\]
We can interpret this as the condition that the current
\[
J_{X}^{m\uba} := \sqrt{-g} T^{mn}\di_n X^{\uba} \lb{scalarCURR}
\]
is conserved.

Let us now turn to the equation of motion for the fermions.  Here we
have to do a bit more work.  Using $\nabla_{m}(Q^{-1} m^{mn})=0$
\cite{HoSeWe97b} we can rewrite \Eq{spinorEOM} as
\[
0
&=& Q^{-1} m^{mn}\nabla_m \gT (1-\gC)\gC_{n}\non
&=&\nabla_m(Q^{-1}m^{mn}\gT(1-\gC)\gC_{n})\\
&=&\frac{1}{\sqrt{-g}}
\di_{m}(\sqrt{-g} Q^{-1}m^{mn}\gT(1-\gC)\gC_{n}).\nn
\]
In the second step we used $m^{mn}\nabla_m ((1-\gC)\gC_n)=0$.  To see
this first note that $\nabla_m \gC_{n}=0$.  This follows from the fact that
the curved $\gC_m$ are constructed from the flat 11-dimensional
$\gC_{\uba}$ as $\gC_{m} = \partial_{m}X^{\uba}\gC_{\uba}$ and the
covariant derivatives $\nabla_m$ are associated with the induced
metric
$g_{mn} =\partial_{m} X^{\uba}\partial_{n} X^{\ubb}\gy_{\uba\ubb}$.
It  remains to show that $m^{mn}\nabla_{m} (\gC_{(h)} \gC_n)=0$.  We
have,  moving to tangent frame,
\[
m^{ab}\nabla_{a} \gC_{(h)} \gC_b &=& \frac13 m^{ab}\nabla_{a}
h^{c_1c_2c_3}\gC_{c_1c_2c_3}\gC_{b}\non
&=&\frac13 m^{ab}\nabla_{a}
h^{c_1c_2c_3} (\gC_{c_1c_2c_3b} + 3 \gy_{c_1b}\gC_{c_2c_3})\non
&=&\frac13 m^{ab}\nabla_{a} h^{c_1c_2c_3} (\frac12\gC_{012345}
\gep_{c_1c_2c_3bc_4c_5}\gC^{c_4c_5}+3
\gy_{c_1b}\gC_{c_2c_3})\lb{consGamma}\\
&=& m^{ab}\nabla_a h_{bc_2c_3}\gC^{c_2c_3}(1-\gC_{012345})\non
&=&0.\nn
\]
We have used the duality relation for the $\gC_{a}$ matrices and in
the last step also the equation of motion for the self-dual
field-strength $m^{ab}\nabla_{a}h_{bcd}=0$ \cite{HoSeWe97a}.
Hence again we can identify a conserved current
\[
J_{\gT}^m := \sqrt{-g} Q^{-1}m^{mn}\gT(1-\gC)\gC_{n}\lb{11current},
\]
where we suppressed the target-space spinor-indices.  Note that due to
the presence of the projector $(1-\gC)$ we only get 16 independent
currents.

Analogously we can rewrite the tensor field equation of
motion \Eq{tensorEOM} as
\[
0=T^{mn} \nabla_{m} H_n{}^{pq}= \nabla_{m} (T^{mn} H_{n}{}^{pq})
= \frac{1}{\sqrt{-g}} \di_{m}(\sqrt{-g}T^{mn} H_{n}{}^{pq}).
\]
In the last step we have used that $T^{mn}
H_{n}{}^{pq}=Q^{-1}G^{mn} H_{n}{}^{pq}$ is totally anti-symmetric in $m,p$
and $q$.  To see this recall that $G^{mn}=(m^2)^{mn}$ and $H_{mnp} =
(m^{-1})_m{}^qh_{qnp}$. This gives
\[
T^{m}{}_{n} H^{npq} = Q^{-1} m^{m}{}_{n}h^{npq} = *\!
H^{mpq},
\]
where $*\! H^{mpq}$ is the dual of $H_{mpq}$
from which the antisymmetry follows. Again we can identify a
conserved current
\[
J_{H}^{mnp} := \sqrt{-g}T^{mq}H_{q}{}^{np} = \sqrt{-g} *\!\! H^{mnp}.
\]

\section{Identifying the Symmetries}

Having found three new conserved currents it is natural to ask what
symmetries the associated charges will generate.  Recall that given
any current satisfying $\di_m J^m=0$ we can define an
associated time-independent charge
\[
Q := \int d^5x J^0
\]
which will generate a symmetry transformation on some field $\Phi$ via
\[
\gd \Phi = \{Q,\Phi\},
\]
where $\{,\}$ denotes the Poisson bracket.

Unfortunately within the covariant approach to the fivebrane used
here we do not have a Lagrangian for the fivebrane fields.  However,
to identify the symmetries it will however be sufficient to keep only
the lowest order terms in  the currents and then use the free-field
Poisson brackets.  We denote these lowest order   currents by the same
symbol the distinction between them and their complete form being
apparent from the context. The lowest order currents are given by
\[
J^{m \uba}_{X} &=& \di^m X^{\uba},\\
J^{mnp}_{H} &=& h^{mnp},\\
J^{m}_{\gT} &=& \gT(1-\gC)\gC^m.
\]
The free-field (equal time) Poisson brackets are given by
\[
\{X^{\uba}(x),\di_{0} X^{\ubb}(x')\} &=& \gy^{\uba\ubb}\gd(x-x'),\non
\{\gT(x),\gT(x')^\dagger\} &=& \gd(x-x'),\lb{pbrack}\\
\{B_{ij},H^{0lk} \} &=& \delta^l_{[i} \delta^k_{j]} \delta (x-x').\nn
\]
Some comments are in order regarding the last equation where we have
used $i,j,\ldots$ to denote purely spatial indices.  Since one does
not have a simple free action for the self-dual gauge field one can
not deduce its Poisson bracket in the standard way.  However, in what
follows it will be sufficient to adopt the above Poisson bracket for
the linearized fields.

Using the explicit Poisson brackets and including constant parameters
one readily checks
\[
\{\xi_{\ubb} \bar{Q}_{X}^{\ubb},X^{\uba}\} = \xi_{\ubb}\int d^5x \{\di^{0}
X^{\ubb },X^{\uba}\} = \xi^{\uba}.
\]
Hence $\bar{Q}_{X}$ generates `shifts' of the scalar fields as expected. For
the spinors we find
\[
\{ \bar{Q}_{\gT}\bar{\gep},\gT\} = \gep(1-\gC),
\]
again constant `shifts'.  Note that the occurrence of the projection
operator implies that only half the Fermions are getting shifted;
these are precisely the dynamical Fermions appearing in the Dirac
equation.

Finally for the
gauge field we have
\[
\{a_{pq}\bar{Q}_{H}^{pq},B_{mn}\} = a_{mn}.
\]
\par
Hence we find that all the fields of the fivebrane undergo shift
 symmetries and therefore the fivebrane itself can be
interpreted as a Goldstone object associated  with  the superalgebra
of  the above generators.  The way the symmetry generators arise
from the equations of motion is also typical of the Goldstone
phenomenon.

\section{The Eleven dimensional Superalgebra}

We now want to determine the algebra of the conserved charges using
the currents with only their lowest order terms.
In this approximation the algebra has only one
non-vanishing Poisson bracket which is given by
\[
\{\bar{Q}_{\gT}\bar{\gep}_1,\bar{Q}_{\gT}\bar{\gep}_2\} &=& 2 \gep_1 \gC_{\uba}
\bar{\gep}_2\bar{Q}_X^{\uba} + 2  \gep_1 \gC_{\uba_1 \ldots \uba_5}
\bar{\gep}_2 Z_5^{\uba_1 \ldots \uba_5}\nonumber\\
&&+2\gep_1 \gC_{\uba_1\uba_2}\bar{\gep}_2 Z_2^{\uba_1
\uba_2}.\lb{11Susy}
\]
Here $Z_5^{\uba_1\ldots \uba_2}= \int d^5x\frac{1}{5!}\gep^{i_1 \ldots
i_5} \di_{i_1} X^{\uba_1} \cdots \di_{i_5} X^{\uba_5}$ is the
five-form central charge associated with the fivebrane and
$Z_2^{\uba_1\uba_2}= \int d^5x h^{0ij} \di_i X^{\uba_1} \di_{j}
X^{\uba_2}$ is the two-form central charge associated with the
membrane.  This is the standard 11-dimensional superalgebra with
five-form and two-form central charges present.  In the fivebrane
context it was first given in \cite{SorTow97} using the Lagrangian
formulation of the fivebrane. We can interpret this result to
say that the presence of the fivebrane generates the central charges
of the eleven dimensional superalgebra.

In deriving \Eq{11Susy} we have used various algebraic properties of the
matrices $\gC_{(h)}$ and $\gC_{(0)}$, these can be found in
\cite{BaLaWe99b}, and also the following Majorana-flip-identity
\[
\psi \gC_{\uba_1 \ldots \uba_m} \bar{\phi} = (-1)^{\frac{m(m+1)}{2}}
\phi \gC_{\uba_1 \ldots \uba_m} \bar{\psi},
\]
which can be derived using the standard properties of the flat eleven
dimensional $\gC_{\uba}$ matrices.

\section{The Worldvolume Superalgebra}

Having reconstructed the eleven dimensional superalgebra from
the equations of motion for the fivebrane we will now make contact
with the (2,0) supersymmetry algebra living on the world-volume of
the five brane.
The fivebrane breaks half of the supersymmetries of the background
superspace and its the worldvolume supersymmetry  can be found
from the eleven dimensional supersymmetry algebra by taking half of the
supersymmetry parameter to vanish, namely
\[
\epsilon ^{\underline{\ga}} = (\epsilon^\ga,0).
\]
Under this truncation  the term that contains
the momentum becomes
\[
\gep_1 \gC_{\uba}\bgep_2 Q_X^{\uba}
= \gep_1 \gC_{\uba}\gC_0\gep^\dagger_2 Q_X^{\uba}
= \gep_1 \gC_m\gC_0\gep^\dagger_2 Q_X^m
= \gep_1 \gC_m\bgep_2 Q_X^m,
\]
where we have used the explicit eleven dimensional $\gC_{\uba}$
matrices given in \cite{GaLaWe98a} to identify nonzero contributions.
The two central charge terms can be reduced in complete analogy and
we  find, omitting the supersymmetry  parameters and using static 
gauge,
\[
\{Q_\ga^i, Q_\gb^j\}= \gy^{ij} (\gc_{m})_{\ga\gb} (P^m)'
                     + (\gc_{m})_{\ga\gb} Z_1^{ij\; m}
		     + (\gc_{m_1m_2m_3})_{\ga\gb} Z_3^{ij\;m_1m_2m_3},
\]
with
\[
(P^m)'          &=& \int d^5x (T^{0m} -\gy^{0m}
),\non
Z_1^{ij\; m} &=& -2 \int d^5x ((\gc_{a'})^{ij} h^{0mn} \di_{n} X^{a'}
                                +\frac{1}{4!} \gc_{a'_1\ldots a'_4}^{ij}
				\gep^{0m n_1\ldots n_4} \di_{n_1} X^{a'_1}
				\cdots \di_{n_4} X^{a'_4}),\non
Z_3^{ij \; m_1m_2m_3} &=& \frac{1}{3!} \int d^5x
(\gc_{a'_1a'_2})^{ij}
\gep^{0 m_1m_2m_3 n_1n_2} \di_{n_1}X^{a'_1}\di_{n_2} X^{a'_2}.				
\]				
This is indeed the six-dimensional $(2,0)$ supersymmetry algebra
\cite{HoLaWe98b} with one-form and three-form central charges present.
The internal symmetry group is
$USp(4) \simeq Spin(5)$, and
$\eta^{ij}$ denotes the associated antisymmetric invariant tensor. The
Weyl spinor indices
$\ga,\gb,\ldots$ run from 1 to 4 as do the internal Spin(5) indices
$i,j,\ldots$.  The $\gc$-matrices arise as building blocks of the
{\em flat} eleven dimensional $\gC^{\uba}$ matrices via
\[
\gC^m = \gd_i^j\left(
\begin{array}{cc}0 & (\gc^m)_{\ga \gb} \\ (\tilde{\gc}^m)^{\ga \gb} &
0
\end{array}
\right)
\quad
\mbox{ and }
\quad
\gC^{a'} = (\gc^{\ga'})_i{}^j\left(
\begin{array}{cc}\gd_\ga{}^\gb & 0\\ 0 & -\delta _{\ \gb}^{\ga}
\end{array}
\right).
\]
We use $m,n,p,\ldots=0,1,\ldots,5$ and $a',b',c',
\ldots=6,7,\ldots,10$ here to denote flat
indices. The basic relations are
\[
\{ \gc^m, \gc^n \} := \gc^m \tilde{\gc}^n + \gc^n
\tilde{\gc}^m=2\gy^{mn},
\]
with $\tilde{\gc}^m = \gc^m$ for $m\neq0$ and $-\tilde{\gc}^0 =
\gc^0=1$. The antisymmetric product is defined as
\[
\gc^{m_1m_2m_3\ldots} := \gc^{[m_1}\tilde{\gc}^{m_2} \gc^{m_3}
\cdots,
\]
and one also has the following duality relation
\[
\gc^{m_1m_2\ldots m_n}= - \frac1{(6-n)!}(-1)^{\frac{n(n+1)}{2}}
\gep^{m_1m_2\ldots m_n m_{n+1}\ldots m_6}\gc_{ m_{n+1}\ldots m_6}.
\]

Note that in the integral representation of the momentum $(P^{m})'$ we
find $T^{mn}-\gy^{mn}$ instead of simply $T^{mn}$; the additional term
appears in the reduction of $Z_5$.  Recall that $T^{mn}$ reduces to
$\gy^{mn}$ for a flat static brane.  However, from the point of view
of the world-volume theory this is a vacuum configuration and has to
have zero energy in a supersymmetric theory.  Hence the tensor that
appears in the $(P^m)'$ above is therefore the natural energy momentum
tensor.  Note also that the one-form central charge in six dimensions
receives contributions from both the five-form and the two-form charge in
eleven dimensions.

We now give the currents that generate the
above worldvolume algebra. Clearly, the current
$J_H^{mnnp}$ associated with the gauge field is   a current
associated with the worldvolume algebra. The
other currents, namely those associated with translations and
supersymmetry transformations,  can be deduced from their eleven
dimensional counter parts by choosing superstatic gauge and
considering only those components that lie in the worldvolume
directions. One then can then verify that  the
resulting current  transform covariantly and are indeed conserved.
For the translations we find that  in static gauge we have
\[
J_X^{mn}=\sqrt{-g} T^{mp}\di_p X^n= \sqrt{-g} T^{mn}.
\]
Which is indeed the conserved current that generates six
dimensional translations \cite{BaLaWe99}.

In superstatic gauge the spinor $\Theta ^{\underline \alpha}$ takes
the form
\[
\gT^{\underline{\ga}} = (\gT^\ga,\gT^{\ga'}).
\]

The linearized six-dimensional supercurrent is given as a
straightforward  reduction of the eleven dimensional one
\Eq{11current} as
\[
(J_{\gT}^m)_\alpha = (\gT (\gc^n \di_n X^{a'}\gc_{a'} - \frac13
h^{n_1n_2n_3} \gc_{n_1n_2n_3}) \gc^{m})_\alpha ,
\]

The full non-linear form of this current can be read off from the
supersymmetry transformation
given  in
\cite{BaLaWe99} and
\cite{HoLaWe98,GaLaWe98a}.
\par
It is straightforward to verify that these currents do indeed lead to
generators that satisfy the above worldvolume superalgebra.

\section{Automorphisms  of  Superalgebras}

Let us consider a supersymmetry algebra of the form
\[
\{Q_\uga ,Q_\ugb \}=Z_{\uga\ugb},\qquad
[Q_\ugc,Z_{\uga\ugb}] = 0,\qquad
[Z_{\uga\ugb}, Z_{\ugc\ugd}]=0,
\]
where $Z_{\uga\ugb}$ are Grassmann even generators labelled by spinor
indices.  It is obviously symmetric in these indices.  Clearly, such
an algebra obeys the generalized super Jacobi identities.  Let us
assume that the generators $Z_{\uga\ugb}$ form the most general
symmetric matrix.  Expanding this matrix out in terms of the relevant
Clifford algebra we find that $Z_{\uga\ugb}$ contains a set of totally
antisymmetric tensorial generators which constitute the central
charges including the momentum generator.  This is the case for both
the eleven dimensional and the six dimensional superalgebras
associated with the fivebrane equations of motion \Eq{scalarEOM} to
\Eq{tensorEOM}.  Clearly in the latter case one must take an
appropriate index range for the indices $\uga , \ugb \ldots$.
 The corresponding
superalgebras in the IIA and IIB theories for which the
$Z_{\uga\ugb}$ form the most general matrix in ten dimensions are well
known.

We now want to consider the automorphism group of such a superalgebra.
Let $R$ be one of the generators of this group.  Being Grassmann even
its commutator with the supercharges must be of the form
\[
[Q_\uga, R]= M_\uga{}^\ugb Q_\ugb.
\]
The super Jacobi identities then imply that
\[
[Z_{\uga \ugb}, R]= M_\uga{}^\ugc Z_{\ugc  \ugb} +
M_\ugb{}^\ugc Z_{\uga \ugc}.
\]
It is straightforward to verify that all the remaining super Jacobi
identities are satisfied provided the matrices $M$ form a
representation of  the Lie algebra generated by the generators
$R$.

Clearly, the maximal automorphism group is $GL(c_d)$ where $c_d$ is
the number  of supercharges. The precise properties of the matrices
under complex conjugation being given by implementing the Majorana
or other properties of the spinorial supercharge on the commutator
relation of the supercharge with the automorphisms.  The
generators may be labelled as $R_\ugc{}^\ugd$ and their commutator
with the supercharges is given by
$[Q_\uga, R_\ugc{}^\ugd]= \gd_\uga{}^\ugd
 Q_\ugc$.  To gain a more familiar set of generators
we may expand $R_\ugc{}^\ugd$ out in terms of the elements of the
enveloping Clifford algebra
\[
R_\ugc{}^\ugd= \sum_p \sum_{\uba_1\ldots \uba_p}
(\ugc^{\uba_1\ldots \uba_p}C^{-1})_\ugc{}^\ugd R_{\uba_1\ldots
\uba_p}\lb{Rdecomp}.
\]
An interesting subalgebra is given by the set of symmetric matrices,
these form Lie algebra of  $Sp(c_d$).  The generators of this
subalgebra are $S_{\ugc\ugd}= R_{\ugc\ugd}+R_{\ugd\ugc}$ where
$R_{\ugc\ugd}= R_{\ugc}{}^{\ugb}(C^{-1})_{\ugb \ugd}$.  Clearly, in
the decomposition of equation \Eq{Rdecomp}  this means keeping only
those terms for which the symmetric matrices enter.  In eleven
dimensions these are the generators
$R_{\uba_1\ldots \uba_p}$ of ranks one, two and five.  The commutator of the
generators $S_{\ugc\ugd}$ with the supercharges is given by
\[
[Q_\uga, S_{\ugc \ugd}]= (C^{-1})_{\uga \ugd}
 Q_\ugc+ (C^{-1})_{\uga \ugc}
 Q_\ugd.
\]
One can also consider the case where $Z_{\uga \ugb}$ is not the
most general symmetric matrix.  For example, when it is of the form
$P^{\underline a}(\gC_{\underline a})_{\uga \ugb}$, implying that we
are dealing with point particles only.  Then the most general
automorphism group is by definition the group Spin($1,d-1$).  These
are the generators $R_{\uba_1\uba_2}$ in the
decomposition of equation
\Eq{Rdecomp}.

Including further central charges one finds a natural generalization
of the spin group that takes into account the presence of branes.  It
would be interesting to compute the relevant spin group
generalizations corresponding to removing various sets of central
charges from the full possible set.

Clearly the generators of the generalized spin group will rotate the
central charges into one another.  As an example let us consider how a
generator of $GL(c_d)$ rotates the central charges.  To be concrete,
let us consider the eleven dimensional supersymmetry algebra of
equation \Eq{11Susy} and consider the effect of the generators
$R_{\uba_1\uba_2\uba_3} \in GL(32)$, not belonging to
Sp(32).  One finds that
\[
[P_{\ubb_1},R^{\uba_1\uba_2\uba_3}]&=& -12
\eta^{\uba_1}_{\ubb_1}Z^{\uba_1\uba_2}, \non
{} [Z_{\ubb_1\ubb_2},R^{\uba_1\uba_2\uba_3}]&=&
2 \eta^{\uba_1\uba_2}_{\ubb_1\ubb_2}P^{\uba_3}
-5!Z^{\uba_1\uba_2\uba_3}_{\ubb_1\ubb_2}, \lb{abstractalg}\\
{}[Z_{\ubb_1\ldots \ubb_5},R^{\uba_1\uba_2\uba_3}]&=&
2 \eta^{\uba_1\uba_2\uba_3}_{\ubb_1\ubb_2\ubb_3}
Z_{\ubb_4\ubb_5}
+3\epsilon^{\uba_1\uba_2\ubc_2\ldots \ubc_5\ubb_1\ldots
\ubb_5}Z^{\uba_3}_{\ \ \ubc_2\ldots \ubc_5}\nn.
\]
Here antisymmetry in $\uba_1\uba_2\uba_3$ is to be understood and
$\eta ^{\uba_1\ldots \uba_n}_{\ubb_1\ldots \ubb_n}=\eta
^{[\uba_1}_{\ubb_1}\ldots
\eta ^{\uba_n]}_{\ubb_n}$.
\par
Since the central charges are the topological charges
for the corresponding branes \cite{Achtow} we must conclude that
the automorphism group must rotate the different branes into each
other.

\section{New Brane Symmetries}

In this section we will show that at least a part of the generalised
spin group, or automorphism group, discussed in the previous section
is realised as a symmetry of the  fivebrane equations of motion. In
particular we will find the associated new current and find the
Poisson bracket relations of its generators with the central
charges.  For simplicity we will keep
working  with a flat background and also ignore terms of higher order
in the  Fermions.

Consider the following current
\[
R_3^{m\uba\ubb\ubc} =  \sqrt{-g}\left(*\!H^{mnp} \nabla_n X^\uba \nabla_p
X^\ubb
X^\ubc+ \frac{1}{2eee}\epsilon^{mnpqrs} \nabla_n X^{\underline a}
 \nabla_p
X^{\underline b}
\nabla_q X^{\underline c} B_{rs}
\right)\lb{R3curr}
\]
where anti-symmetry in ${\uba\ubb\ubc}$ is assumed. To check that it
is conserved recall the following facts
\[
\nabla_m *\!H^{mnp}=0,\qquad H_{mnp} = 3\nabla_{[m} B_{np]}.
\]
Note that \Eq{R3curr} can be rewritten as 
\[
R_3^{m\uba\ubb\ubc}= \di_n \left(\frac12 \sqrt{-g} 
\gep^{mnpqrs}\nabla_{p}X^{\uba} \nabla_q X^{\ubb} X^{\ubc} 
B_{rs}\right),
\]
from which conservation is obvious.

Given this additional current we can use the same procedures as
earlier and compute various Poisson brackets at lowest order.  To make
our notation less cumbersome we will use the same symbol to denote the
associated conserved charge.

We now calculate  the commutators of the $R_3^{\uba\ubb\ubc}$ with the
central terms. Using the Poisson brackets of equations \Eq{pbrack} we
find that
$$[P_{\ubb_1},R^{\uba_1\uba_2\uba_3}]=
\eta^{\uba_1}_{\ubb_1}Z^{\uba_2\uba_3}, \ \
[Z_{\ubb_1\ubb_2},R_{\uba_1\uba_2\uba_3}]=
\frac12Z_{\uba_1\uba_2\uba_3\ubb_1\ubb_2}, \ \  [Z_{\ubb_1\ldots
\ubb_5},R_{\uba_1\uba_2\uba_3}]= 0.
$$
Up to normalisation this is a contraction of the algebra of equation
\Eq{abstractalg} obtained by scaling the generators as follows
\[
P^{\uba}\to P^{\uba}, \ \ R_{\uba_1\uba_2\uba_3} \to
\frac1{e} R_{\uba_1\uba_2\uba_3}
\ Z^{\uba_1\uba_2}\to \frac{1}{e} Z^{\uba_1\uba_2},\ \
Z^{\uba_1\ldots \uba_5}\to \frac{1}{e^2} Z^{\uba_1\ldots \uba_5}.
\]
and letting $e\to 0$.  It is conceivable that the contraction is the
effect of our approximation and that by carrying out the calculation
for the full theory one may find the commutation relations of the
full automorphism group.

\section{Conclusions}

We have shown, in the context of the covariant equation of motion,
that the fivebrane possesses symmetries that lead to the worldvolume
supersymmetry algebra and  the eleven dimensional supersymmetry
algebra. We have explicitly identified the corresponding currents
by writing the equations of motion as total derivatives. In addition
to the obvious currents for translations and supersymmetry we find
that there exists a conserved  current associated with the second rank
tensor gauge field  of the fivebrane. The generators of these
currents lead to shift symmetries for all the fields of the fivebrane
so allowing us to identify them as Goldstone fields for these
symmetries.

Motivated by the observation that  both eleven dimensional and
worldvolume superalgebras associated with the fivebrane have central
charges that constitute  the most general symmetric matrix on the right
hand-side of the anti-commutator of two supercharges we show that  the
automorphism group of such a generic superalgebra is $GL(c_d)$ where
$c_d$  is the number of supercharges. This is the natural
generalisation of the spin group which occurs in point particle
theories to a theory, such as M theory,  that possess branes as well
as point particles.  Such an automorphism group must rotate the
different branes into one another and so is a generalisation of
the previously known duality symmetries.

We show, at lowest order, that part of this automorphism group is
realised as a new symmetry of the fivebrane equations of motion.  At
first sight, it may appear as a puzzle that this new brane rotating
symmetry can be seen solely within the context of the fivebrane. 
However, the fivebrane through its equations of motion can see many of
the effects in the eleven dimensional superspace including the
presence of twobranes.  Indeed, the twobranes which intersect
fivebranes induce strings on the fivebrane whose charge is measured by
the gauge field of the fivebrane.  Examining the new current we find
that the gauge field plays the central role in its construction
consistent with its brane rotating property.

As such, it is natural to speculate that the $GL(32)$ automorphism
group of the eleven dimensional supersymmetry algebra, or a subgroup
of it, is a symmetry of M theory when properly formulated. 
Incorporating this symmetry would inevitably lead to a formulation in
which the fivebrane and twobrane appeared on an equal footing.  Since
the symmetry rotates the twobrane and the fivebrane into each other
one may wonder if it is a hidden symmetry of the eleven dimensional
supergravity that rotates the corresponding brane solutions into one
another.  In the quantum theory the topological charges of the branes
obey a generalisation of the Dirac quantization condition and so in the
quantum theory one should only attempt to incorporate the subalgebra
of the automorphism group that preserves this relation.

The new symmetries we have found are not just restricted to the branes
of M theory, but by reducing the fivebrane to the D4 brane in ten
dimensions and then using T duality we can extend it to be relevant to
all the branes in ten dimensions.

\medskip\noindent {\bf Acknowledgment:} We would like to thank Neil 
Lambert for helpful discussions. While this work was being
written up we learned from Garry Gibbons that he, Jerome Gauntlett,
Chris Hull and Paul Townsend were preparing a paper in which the
automorphism group of superalgebras is considered within the context
of BPS states and, in particular, those that preserve ${3/4}$
supersymmetry.

\end{document}